\documentclass[prb,aps,twocolumn,floatfix,citeautoscript,noshowpacs]{revtex4}
\usepackage{graphicx,rotating,subfigure,amsmath,amsfonts,amssymb,delarray,color}
\usepackage{dsfont}
\usepackage[T1]{fontenc}
\usepackage{multirow}
\usepackage{comment}
\renewcommand{\vec}[1]{\boldsymbol #1}
\newcommand{\e}{\text{e}}

\def\12{\frac{1}{2}}

\begin{document}
\title{Does a distinct quasi many-body localized phase exist?\\
A numerical study of a translationally invariant system in the thermodynamic limit}
\author{J.~Sirker}
\affiliation{Department of Physics and Astronomy, University of Manitoba, Winnipeg R3T 2N2, Canada}
\date{\today}

\begin{abstract}
We consider a quench in an infinite spin ladder describing a system
with two species of bosons in the limit of strong interactions. If the
heavy bosonic species has infinite mass the model becomes a spin chain
with quenched binary disorder which shows true Anderson localization
(AL) or many-body localization (MBL). For finite hopping amplitude
$J'$ of the heavy particles, on the other hand, we find an exponential
polarization decay with a relaxation rate which depends monotonically
on $J'$. Furthermore, the entanglement entropy changes from a constant
(AL) or logarithmic (MBL) scaling in time $t$ for $J'=0$ to a
sub-ballistic power-law, $S_{\textrm{ent}}\sim t^\alpha$ with $\alpha<
1$, for finite $J'$. We do not find a distinct regime in time where
the dynamics for $J'\neq 0$ shows the characteristics of an MBL
phase. Instead, we discover a time regime with distinct dephasing and
entanglement times, leading to dynamics in this regime which is
different {\it both} from a localized and from a fully ergodic phase.
\end{abstract}

\maketitle

\section{Introduction}
The wave function of a single quantum particle in one dimension always
becomes localized in a disorder potential
\cite{Anderson58,AbrahamsAnderson,AndersonLocalization}. Anderson localization (AL)
of non-interacting quantum particles has been observed, for example,
in ultracold gases
\cite{RoatiDerrico,BillyJosse}. Because AL is at its heart an interference 
phenomenon it has also been studied for classical waves
\cite{SoundLocalization}. 

In recent years, the interplay of disorder and interactions in
many-body quantum systems has attracted renewed interest
\cite{AleinerAltshuler}. Focussing on the one-dimensional case, a 
number of analytical and numerical studies have shown that, under
certain conditions, a many-body localized (MBL) phase can occur
breaking the ergodicity of the system
\cite{PalHuse,Imbrie2016,AndraschkoEnssSirker,EnssAndraschkoSirker,Luitz1,Luitz2,VoskHusePRX,PotterVasseurPRX,NandkishoreHuse,AltmanVoskReview}. 
A key signature of the fully many-body localized phase is the
logarithmic spreading of entanglement entropy,
$S_{\textrm{ent}}\sim\ln t$, under unitary time evolution starting
from an unentangled initial state
\cite{ZnidaricProsen,BardarsonPollmann}. In contrast,
$S_{\textrm{ent}}\sim\mbox{const}$ at long times $t$ in an AL
phase. On the other hand, both in the AL and in the MBL phase the
system retains memory of its initial state. This has been demonstrated
in a cold gas experiment where an interacting quasi one-dimensional
system was prepared in an initial charge-density wave state with the
order parameter being stable over time \cite{BlochMBL}.

Another interesting question which has attracted attention recently
but has been much less explored so far is whether translationally
invariant many-body systems can also have {\it dynamically created}
localized phases or at least extended time regimes where the dynamics
appears to be localized ('quasi MBL') before thermalization ultimately
sets in
\cite{SchiulazMueller,SchiulazSilva,CarleoBecca,GroverFisher,DeRoeckHuveneers,YaoLaumann,MichailidisZnidaric}. 
One of the defining properties of quasi-MBL according to
Ref.~\onlinecite{YaoLaumann} where this terminology was introduced is
that ``The entanglement dynamics are consistent with MBL-type growth
at short and intermediate times, but ultimately give way to anomalous
diffusion.'' In contrast to disorder-induced MBL where localization
occurs despite interactions, the idea here is that sufficiently strong
interactions might actually induce true localization or quasi
MBL. Studies of spin chains with potential disorder in fact do provide
evidence that very strong interactions can support localization. This
can lead to a reentrant behavior where the system transitions at fixed
disorder from Anderson $\to$ MBL $\to$ ergodic $\to$ MBL with
increasing interaction strengths
\cite{BarLevCohen,EnssAndraschkoSirker,KudoDeguchi18}. On the other hand, translational 
invariance requires that for a finite-size system any finite
wavelength inhomogeneity in the initial state decays to zero in the
infinite time average 
(see Eq.~(4) in Ref.~\onlinecite{YaoLaumann}). True localization in a
translationally invariant system therefore has to be understood as a
divergence of the decay time for initial inhomogeneities with system
size
\cite{YaoLaumann}. In the case of quasi MBL, on the other hand, the
question is if an extended regime in time exists in which the decay of
inhomogeneities is anomalously slow and the entanglement dynamics
consistent with the logarithmic scaling in time expected in an MBL
phase.

Numerical studies have provided evidence that strong interactions can
stabilize inhomogeneous initial states resulting in a very slow
thermalization process
\cite{KollathLauchli,BarmettlerPunk,BarmettlerPunk2,EnssSirker}. In 
Ref.~\onlinecite{CarleoBecca} the possibility was raised that strong
interactions might even induce a delocalization-localization
transition for such inhomogeneous initial states. Even more relevant
for the following discussion are models consisting of a light and a
heavy species which interact with each other
\cite{SchiulazMueller,GroverFisher,SchiulazSilva,YaoLaumann}. Here the idea is that 
the heavy particles might create an effective disorder potential for
the faster light particles driving potentially a transition where the
light particles localize or, alternatively, leading to an extended
regime in time where the light particles appear localized although the
system becomes ultimately ergodic at very long times ('quasi MBL').

It is well known that both AL and MBL can occur in translationally
invariant systems with two particle species if one of the species is
static. Here the static particle species creates a discrete disorder
potential for the mobile one
\cite{ParedesVerstraete}. In particular, the case of spinless fermions
with a nearest-neighbor density-density interaction---which is
equivalent to a $s=1/2$ XXZ spin chain by Jordan-Wigner
transformation---in an effective binary disorder potential has been
studied in detail \cite{AndraschkoEnssSirker,EnssAndraschkoSirker}.

In this paper we will extend these studies to the case where the
second heavy species becomes mobile. Our goal is to study a specific
microscopic model which does show an MBL phase in the static case and
ask whether the MBL phase survives for small hopping amplitudes $J'$
(in the sense of a diverging decay time for initial inhomogeneities)
or at least shows a distinct time regime where MBL characteristics
remain present. We will use an unbiased density-matrix renormalization
group algorithm (see Sec.~\ref{Model} for details)
to obtain numerically exact results for the time evolution of the
system after a quantum quench directly in the thermodynamic limit. We
will concentrate on interactions between the two particle species
which are effectively infinite in the time regime studied. The
advantages of the numerical study presented here as compared to
previous exact diagonalization and perturbative studies are that
numerically exact results in the thermodynamic limit are obtained for
a model which can be, in principle, realized in experiment.

Our paper is organized as follows. In Sec.~\ref{Model} we introduce
the considered $s=1/2$ spin ladder and discuss the numerical algorithm
to calculate the dynamics of observables following the quench. We
present results for the case where both legs of the spin $s=1/2$
ladder are of XX type in Sec.~\ref{Anderson}. Results where the spins
in one leg have an XXZ-type interaction are then discussed in
Sec.~\ref{XXZ}. We summarize and conclude in Sec.~\ref{Concl}.

\section{Model and Methods}
\label{Model}
We start from a bosonic Hubbard model with two species $a$ and $b$
described by the Hamiltonian
\begin{eqnarray}
 \label{Hbos}
H &=& -\frac{J}{2}\sum_j \left(a^\dagger_j a_{j+1} +h.c\right)-\frac{J'}{2}\sum_j \left(b^\dagger_j b_{j+1} +h.c.\right) \nonumber \\
 &+&\sum_j \left[V_a n^a_jn^a_{j+1}+U_a n^a_j(n^a_j-1) + (a\leftrightarrow b)\right] \nonumber \\
&+& D\sum_j n_j^a n_j^b.
\end{eqnarray}
Here $J$ and $J'$ are hopping amplitudes and $U_{a,b}$ the onsite
repulsive interactions for the $a,b$ particles respectively. $D$ denotes
the intraspecies interactions and we also include interspecies
nearest-neighbor terms $V_{a,b}$.

We prepare the system of $b$ particles in a Fock state $|\Psi_b\rangle
= \sum_{n^b_1,\cdots,n^b_M}
\alpha_{n^b_1,\cdots,n^b_M} |n^b_1\,\cdots\, n^b_M\rangle$ where
$n^b_j$ is the occupation number for the $b$ particles at lattice site
$j$ \cite{ParedesVerstraete}. The $a$ particles, on the other hand,
are prepared in an initial product state. If we now time evolve the
system in the limit $J'\to 0$ then the intraspecies interaction term
turns into an effective discrete random potential for the mobile $a$
particles with strengths $Dn_j^a n_j^b \to D P_j^b n_j^a$ where $P_j^b\in
\{0,1,2,\cdots\}$ and the probability distribution is determined by $|\alpha_{n^b_1,\cdots,n^b_M}|^2$. 
In the limit $U_b\to\infty$ where $n_j^b\in\{0,1\}$, in particular, a
binary disorder potential is realized which has been shown to lead to
many-body localization of the $a$ particles at sufficiently large $D$
\cite{AndraschkoEnssSirker,EnssAndraschkoSirker}.

Here we are interested in investigating the case $J'\neq 0$ with $J'<
J$. The question then is if the heavy $b$ particles can still serve as
an effective dynamic binary disorder potential for the light $a$
particles. In order to reduce the number of parameters in the model
and to obtain numerical results for times much longer than $1/J'$ we
concentrate on the hardcore boson case $U_{a,b}\to\infty$ with
$V_b=0$. In this limit the bosonic model
\eqref{Hbos} can be mapped onto the following spin model
\begin{eqnarray}
\label{Ham}
H &=& J\sum_j \left[\frac{1}{2}\left(S^+_jS^-_{j+1} + S^-_jS^+_{j+1}\right) +\Delta S^z_j S^z_{j+1}\right] \nonumber \\
 &+& J' \sum_j \frac{1}{2}\left(\sigma^+_j\sigma^-_{j+1} + \sigma^-_j \sigma^+_{j+1}\right) + D\sum_j S^z_j \sigma^z_j  
\end{eqnarray}
with $J\Delta = V_a$ and spin-$1/2$ operators $\vec{S}_j$ and
$\vec{\sigma}_j$ representing the two species of hardcore bosons. 
For $\Delta=0$ this is {\it exactly the same model} which has been
considered in Ref.~\onlinecite{YaoLaumann} for small clusters of up to
$L=8$ sites.

Here we want to study this model in the {\it thermodynamic limit} for
$\Delta\in [0,1]$. As initial state for the spin ladder we consider
$|\Psi\rangle = |N\rangle_S \otimes |\infty\rangle_\sigma$ where
$|N\rangle = |\uparrow\downarrow
\uparrow\downarrow \cdots\rangle_S$ is the N\'eel state and
$|\infty\rangle_\sigma =\bigotimes_j
\frac{1}{\sqrt{2}} (|+\rangle + |-\rangle)_j$ is the
product state corresponding to an equal superposition of all
arrangements of spins $\sigma^z_j=+,-$. Importantly, the time
evolution starting from the product state $|\infty\rangle_\sigma$ for
$J'=0$ then gives an exact average over all possible effective binary
magnetic field configurations, $D\sigma^z_j\to h_j^{\textrm{eff}}=\pm
D/2$
\cite{ParedesVerstraete,AndraschkoEnssSirker,EnssAndraschkoSirker,TangIyer}. 
Note that for $J'\neq 0$ this setup is no longer equivalent to
averaging over the dynamics starting from each possible configuration
for the slow particles separately. However, setting up the state
$|\infty\rangle_\sigma$ is in principle possible---for example in
experiments on ultracold bosonic quantum gases described by the
effective Hamiltonian
\eqref{Hbos}---and is an interesting starting point because it
does include the case $J'\to 0$ where an exact disorder average is
obtained and a many-body localized phase is known to exist; see
Ref.~\onlinecite{EnssAndraschkoSirker} for a phase diagram of the
model for $J'=0$ as a function of $D$ and $\Delta$.

We want to stress that the dynamics at times $t> 1/J'$ is not expected
to be qualitatively different for different non-trivial initial
states. In Appendix \ref{App_A} we indeed show by exact
diagonalizations that the dynamics starting from the chosen initial
state is {\it qualitatively the same} as the dynamics obtained by
averaging over initial product states as has been done in
Ref.~\onlinecite{YaoLaumann}.

We study the quench dynamics in the spin ladder \eqref{Ham} using the
light cone renormalization group (LCRG) algorithm
\cite{EnssSirker,AndraschkoEnssSirker,EnssAndraschkoSirker}. This
algorithm makes use of the fact that even in the clean ergodic case,
information and correlations for a generic Hamlitonian with
short-range interactions only spread through the lattice at a {\it
finite} (Lieb-Robinson) velocity $v_{LR}$ \cite{LiebRobinson}. Using a
Trotter-Suzuki decomposition for the time evolution operator the
one-dimensional quantum model is first mapped onto a two-dimensional
classical model. It then suffices to consider a finite light cone with
Trotter velocity $v_T\gg v_{LR}$ to be effectively in the
thermodynamic limit. The light cone is extended and then truncated
using a density-matrix renormalization group (DMRG) procedure
propagating the initial state forward in time. The transfer matrices
used to expand the light cone have dimension $4\chi\times 4\chi$ for
the spin ladder and we keep up to $\chi=20\, 000$ states requiring up
to $450$ GB of RAM. We adapt the number of states $\chi$ such that the
truncation error is less than $10^{-10}$ at small and intermediate
times and at most $10^{-7}$ at the longest simulation times shown.

We will concentrate here on two observables as a function of time $t$
after the quench: the staggered magnetization $m_s(t)=\sum_j (-1)^j
\langle S^z_j\rangle$ and the entanglement entropy $S_{\textrm{ent}}(t)$ 
when cutting the whole system---consisting of slow and fast
particles---in two semi-infinite halfs. The asymptotics of these two
observables at long times allows to distinguish between different
phases of the Hamiltonian \eqref{Ham}: (i) If the system is in an AL
phase then $m_s(t\to\infty)\neq 0$ and $S_{\textrm{ent}}(t\to
\infty)\sim
\textrm{const}$. (ii) If the system is ergodic then $m_s(t\to\infty)=0$. 
The entanglement entropy in the clean case without disorder grows
linearly in time. (iii) Finally in the MBL phase, $m_s(t\to \infty)\neq 0$
while asymptotically $S_{\textrm{ent}}\sim \ln t$.

In the following two sections we present our numerical results for the
model \eqref{Ham} in the thermodynamic limit. The analysis of the
scaling behavior is based on data for times $t\lesssim 10/J'$. The
time scale $t\sim D/J^2$ set by the intraparticle interaction $D$---
which is very large in our simulations and effectively infinite in the
time intervals studied---is out of reach and plays no role. The
obtained scaling of $m_s(t)$ and $S_{\textrm{ent}}(t)$ is valid in the
regime $1/J<t<D/J^2$ while fully ergodic behavior is expected to set
in at the longest time scale $t > D/(J')^2$.

\section{The XX case}
\label{Anderson}
For $\Delta=0$ the interactions on both legs of the spin ladder
\eqref{Ham} are of XX type. For $J'=0$ we can map the model by a
Jordan-Wigner to non-interacting spinless fermions subject to a binary
disorder potential of strength $D$. For $D\neq 0$ we thus expect the
model to be in an AL phase. Interestingly, the model becomes
interacting by allowing for a finite hopping $J'$. I.e., we expect
that the hopping drives a transition from an AL phase of
non-interacting fermions to an interacting phase.

To investigate this transition, we concentrate on the case $D\gg J >
J'>0$. Then there is an energy cost $D$ associated with flipping two
neighboring spins $\vec{S}_j$, $|\uparrow\downarrow\rangle \to
|\downarrow\uparrow\rangle$, if the spins $\vec{\sigma}_j$ are
antiparallel as well. One might therefore expect that second order
processes are required for the heavy species to become mobile by
either moving on their own with effective hopping amplitude
$J'_{\textrm{eff}}=(J')^2/D$ or by moving together with the lighter
species with effective hopping amplitude $J_c = JJ'/D$. This then
would give rise to time scales $t'_{\textrm{eff}}=1/J'_{\textrm{eff}}$ and
$t_c =1/J_c$ respectively. 
Note that these timescales are effectively infinite for the time
intervals studied numerically in the following because we set
$D/J=4000$.

If these are the only relevant time scales, then the staggered
magnetization $m_s(t)$ will remain essentially constant for times
$t<\min(t'_{\textrm{eff}},t_c)$ and we then might indeed expect a time
regime where the system appears localized. In the following we will,
however, show that this is not the case and that the dynamics is
actually more complicated: neighboring clusters where spins
$\vec{\sigma}_j$ are parallel can act as a 'bath' and a decay of
$m_s(t)$ already sets in at much smaller time scales.

Let us start with the simplest case of $J'=0$. Then the chain
effectively separates into finite segments with equal potentials $\pm
D$
\cite{AndraschkoEnssSirker}. The probability to find a segment of
length $\ell$ is given by $p_\ell =\ell/2^{\ell+1}$ and the
magnetization becomes
\begin{equation}
\label{segments}
m_s(t\ll D/J^2) = \sum_\ell p_\ell m_s^\ell(t) 
\end{equation}
where $m_s^\ell$ is the staggered magnetization for a segment of
length $\ell$. At times $1/J\ll t\ll D/J^2$ the staggered
magnetization then oscillates around $1/6$ and only the odd clusters
contribute to the non-zero average \cite{AndraschkoEnssSirker}. For
$J'=0$ there is thus no decay at times $t\ll D/J^2$. While this simple
picture breaks down at times $t>D/J^2$, $m_s(t)$ will remain finite
because any amount of potential disorder will lead to Anderson
localization in one dimension.

\begin{figure}[!ht]
\includegraphics*[width=0.99\linewidth]{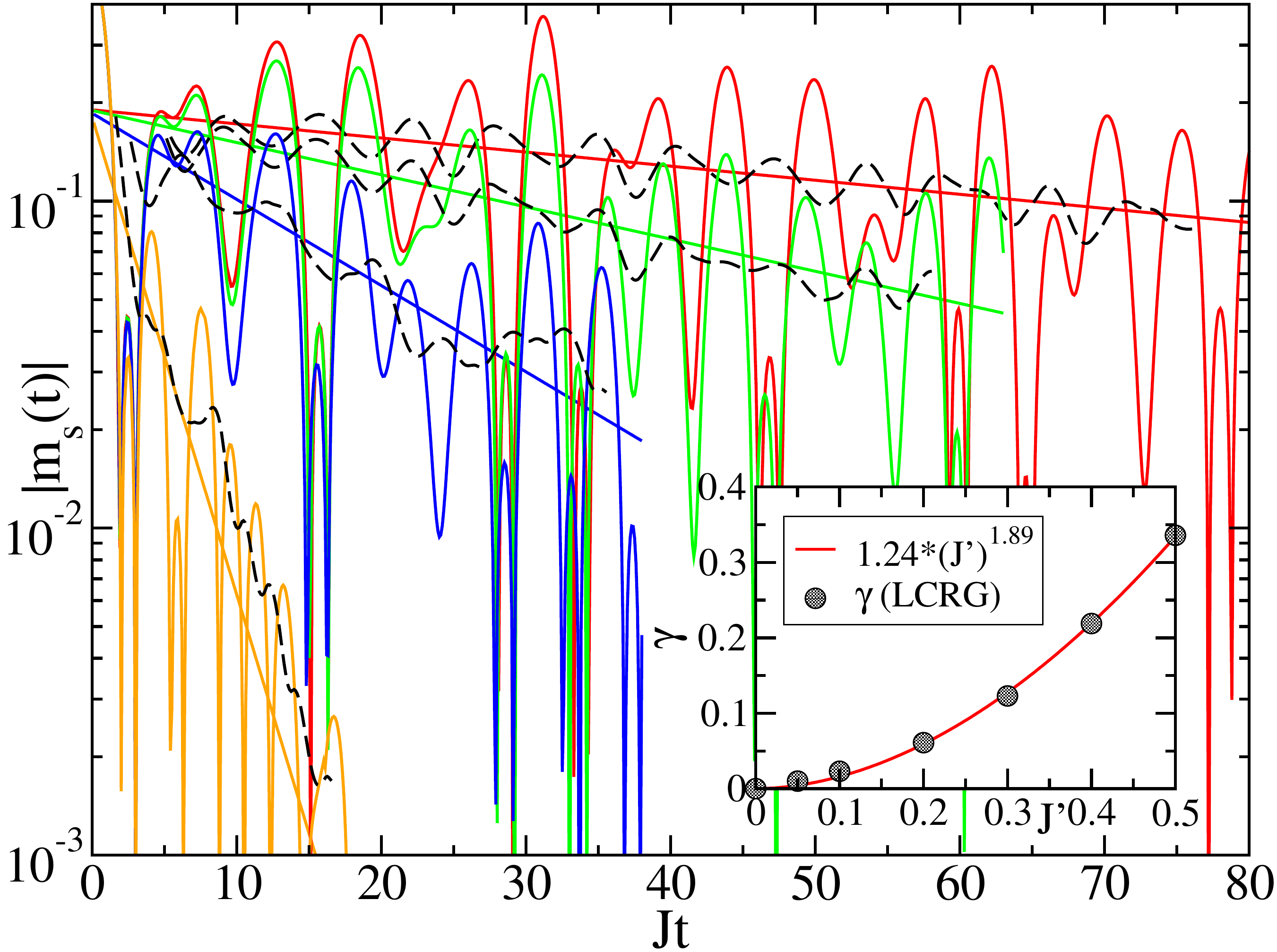}
\caption{Main: Staggered magnetization $m_s(t)$ at $\Delta=0$ and $D=4000$. Shown are LCRG results (solid curves), exponential fits (solid lines), and running averages (dashed lines, guide to the eye) for $J'=0.05,0.1,0.2,0.5$ (from top to bottom). Inset: Relaxation rate $\gamma$ extracted from the exponential fits as a function of hopping amplitude $J'$ (symbols) and a power-law fit (solid curve).}
\label{Fig1}
\end{figure}

\begin{figure}[!ht]
\begin{center}
\includegraphics*[width=0.99\linewidth]{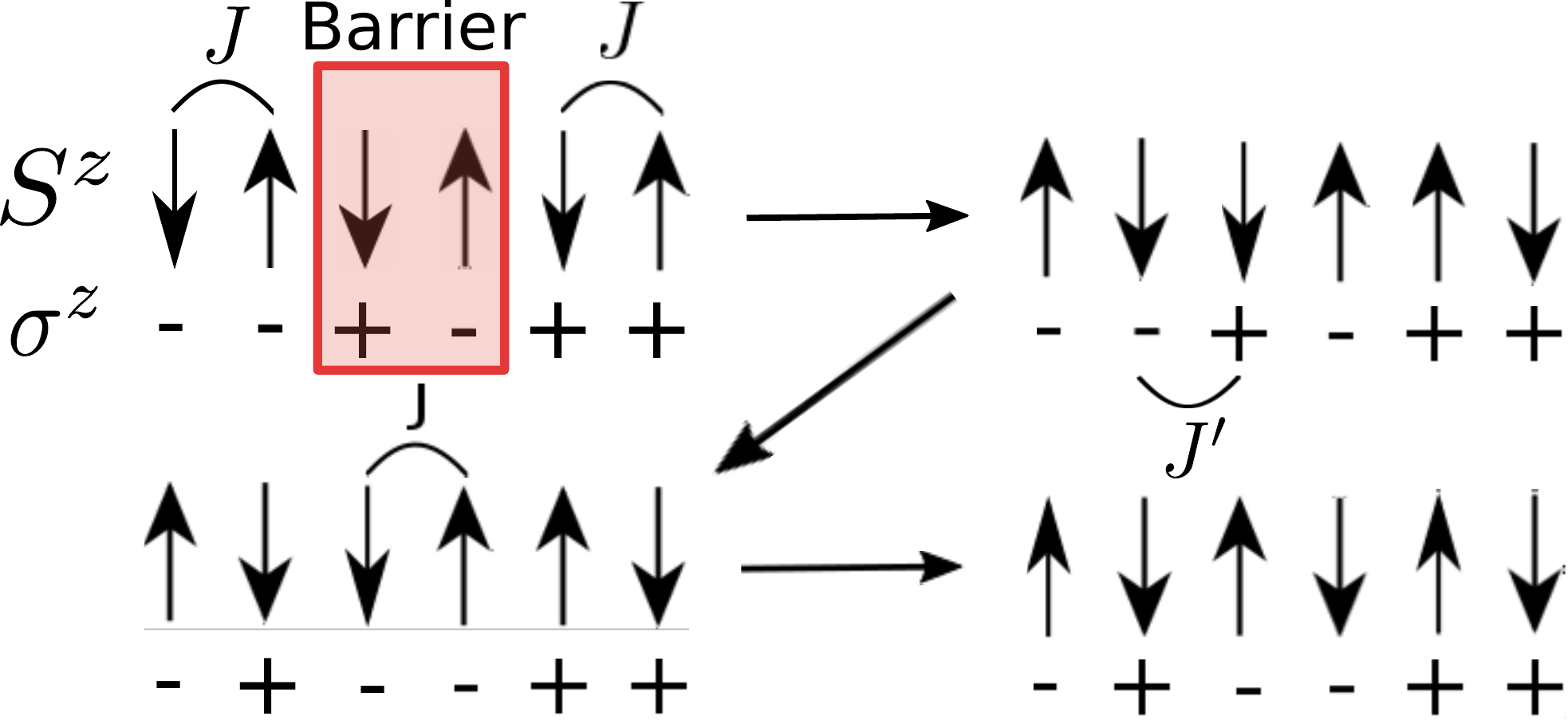}
\end{center}
\caption{From left to right, top to bottom: In the initial state the spins $S^z_j$ (arrows) are in a N\'eel state, the spins $\sigma^z_j$ (+/-) in an equal superposition of all configurations. A $'+-'$ arrangement creates a barrier for spin flips. A three-step process, however, allows to flip all the spins $S^z_j$ without involving the large enery scale $D$.}
\label{Fig2}
\end{figure}

Turning on a weak hopping for the heavy species we would expect that
$m_s(t)$ remains largely unchanged for times $t<\min(t',t_c)$ if the
discussed second order processes involving the large energy scale $D$
are the only ones which can lead to a relaxation. This is, however,
not the case as can clearly be seen from the unbiased numerical data
shown in Fig.~\ref{Fig1}.
Instead, the data are consistent with an exponential decay $m_s(t)\sim
\exp(-\gamma t)$ with a relaxation rate $\gamma\sim A\cdot (J')^{\beta}$
with $\beta=1.89\pm 0.1$ and an amplitude $A\approx 1.2$ which is three
orders of magnitude larger than expected if the responsible processes
would involve the large energy scale $D$ (see inset of
Fig.~\ref{Fig1}). This suggests that there is another mechanism which
leads to a relaxation of $m_s(t)$. This mechanism is depicted in Fig.~\ref{Fig2}.
If we think of flips of neighboring spins as gates applied at a
particular time step in a Trotter-Suzuki decomposition of the time
evolution operator, then we see that three time steps are required to
flip all the spins $S^z_j$ for the single barrier case shown in
Fig.~\ref{Fig2}. Similar processes also exist for larger barriers. We
also note that this process {\it is generic and is important for the
dynamics independent of the initial state}. If $N$ is the number of
consecutive $'+-'$ configurations of $\sigma^z_j$ (number of barriers)
then we find that $N+2$ time steps are required to overcome the
barrier. This implies a dephasing time $\tau\sim f(J',D)N$ across such
barriers where $f(J',D)$ is a function depending on the hopping $J'$
and the intra-particle interaction $D$. We want to stress once more
that we are concerned with the regime $D\gg J>J'$ where the processes
depicted in Fig.~\ref{Fig2} set the relevant time scale for dephasing
and not $J'_{\textrm{eff}}$ and $J_c$ which involve the numerically
infinite energy scale $D$. In this case, the staggered magnetization
at time $t$ is dominated by clusters of size $N\geq N_0=f^{-1}(J',D)t$
which occur with probability $P(N)$ and have remained static. This
implies a scaling\cite{EnssAndraschkoSirker}
\begin{equation}
\label{mstaggt}
m_s(t)\sim \int_{N_0}^\infty P(N)dN\sim \int_{f^{-1}(J',D)t}^\infty\frac{dN}{2^{2N}} \sim \e^{-\frac{t}{f(J',D)}}.
\end{equation}
I.e., these considerations imply that {\it as soon as $J'\neq 0$ there is a
finite decay rate $\gamma\sim f^{-1}(J',D)$.} 

It is also worth comparing this to the case of small quenched binary
disorder where the system is close to the MBL phase but still on the
ergodic side. In this case the time to overcome barriers between
thermalizing clusters scales as $\tau\sim\e^N$. Using again
Eq.~\eqref{mstaggt} then leads to a much slower power-law decay of the
staggered magnetization
\cite{EnssAndraschkoSirker}. The system with mobile heavy particles 
is thus {\it qualitatively different} from a system with quenched
disorder close to the ergodic-MBL phase transition.

If the spin inhomogeneity can decay, then we also expect that
information can spread beyond the finite segments which form for
$J'=0$. We therefore investigate the entanglement entropies between
the two halfs of the infinite ladder next.
\begin{figure}[!ht]
\begin{center}
\includegraphics*[width=0.99\linewidth]{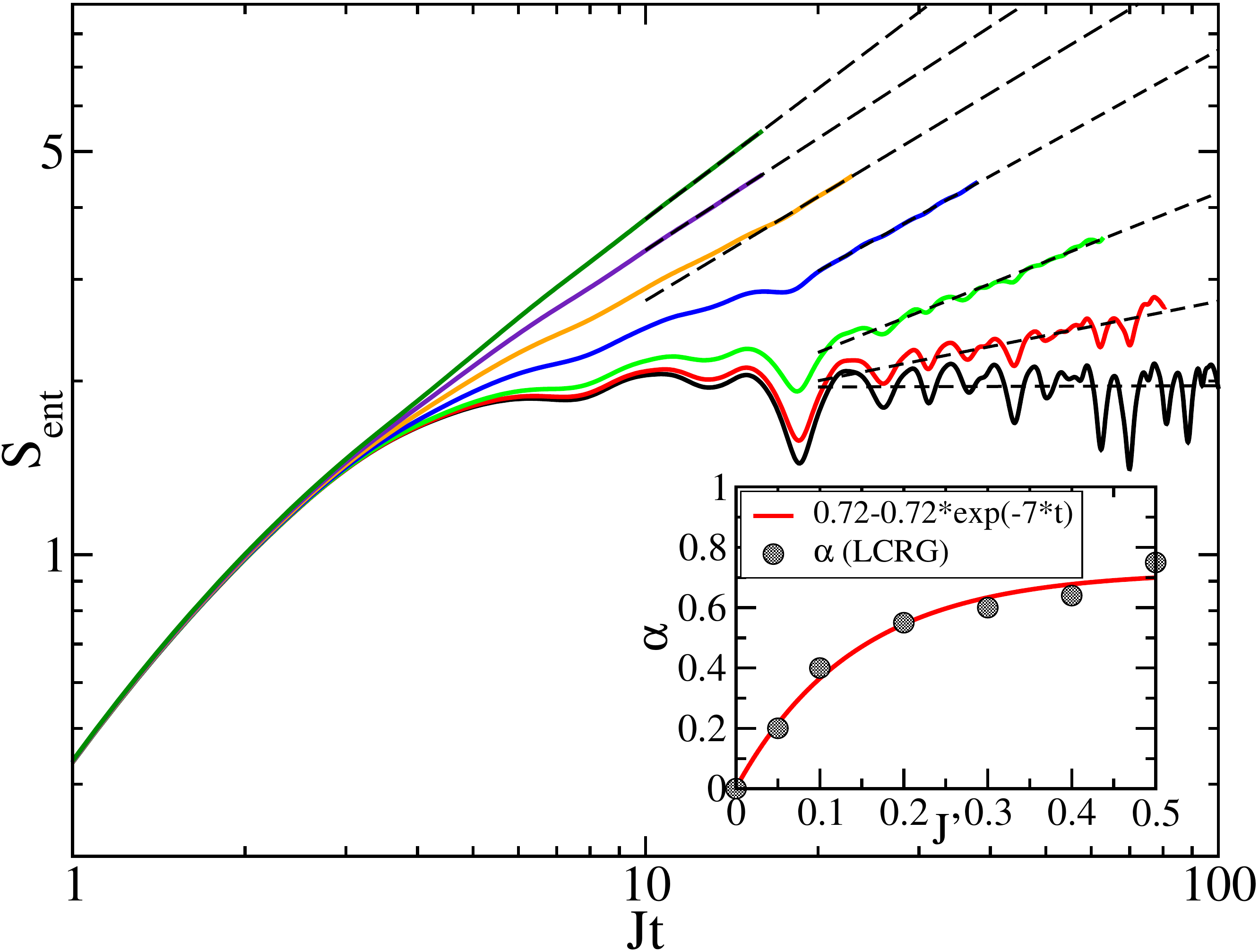}
\end{center}
\caption{Main: Entanglement entropy at $\Delta=0$ and $D=4000$. Shown are LCRG results (solid curves) for $J'=0,0.05,0.1,0.2,\cdots,0.5$ (bottom to top) and power-law fits (dashed lines). Inset: Power-law exponent $\alpha$ as a function of hopping amplitude $J'$.}
\label{Fig3}
\end{figure}
The data shown in Fig.~\ref{Fig3} are consistent with a power-law
scaling, $S_{\textrm{ent}}(t)\sim t^\alpha$, with an exponent $\alpha$
which changes monotonically as a function of the hopping amplitude
$J'$ (see inset of Fig.~\ref{Fig3}). For $J'\sim J$ and $D=4000$ the
exponent seems to saturate to a value $S_{\textrm{ent}}\approx
1/\sqrt{2}\approx 0.71$.

It is important to note that the process depicted in Fig.~\ref{Fig2}
is a process which allows for the dephasing of the staggered
magnetization inside barriers. Thus $\tau\sim N$ is a dephasing
time. It is {\it not} the time required to transport information
across a barrier of size $N$. If this would be the case then
$S_{\textrm{ent}}= vt$ and only the velocity $v$ would change as a
function of $J'$. That the dephasing time $\tau$ and the entanglement
time $\tau_{\textrm{ent}}$ are different can be understood already in
a classical picture: Imagine e.g.~the left spin in the initial
configuration shown in Fig.~\ref{Fig2} as being distinguishable from
the other down spins. Trying to move this spin from the left to the
right end of the segment across the barrier, it becomes obvious that
many more time steps are required than for simply flipping all spins
of the segment. Thus we expect $\tau_{\textrm{ent}} \gg\tau$ as well
as a different scaling with the size of the barrier.

The entanglement entropy is proportional to the entangled region. Thus
if $\tau_{\textrm{ent}}\sim
\e^N$ then $S_{\textrm{ent}}\sim \ln t$ as in the MBL phase. If, on the other hand, 
$\tau_{\textrm{ent}}\sim N^{1/\alpha}$ with $\alpha < 1$ then this
leads to the observed sub-ballistic increase, $S_{\textrm{ent}}\sim
t^\alpha$. Because of the very strong coupling $D=4000$ between the
legs, information spreading is sub-ballistic for $t<D/J^2$ even if
$J'\sim J$. We expect that this sub-ballistic behavior will give way
to a ballistic spreading at times $t>\min(t'_{\textrm{eff}},t_c)>
D/J^2$. Studying the latter regime is outside the scope of
this article.

\begin{figure}[!ht]
\includegraphics*[width=0.99\linewidth]{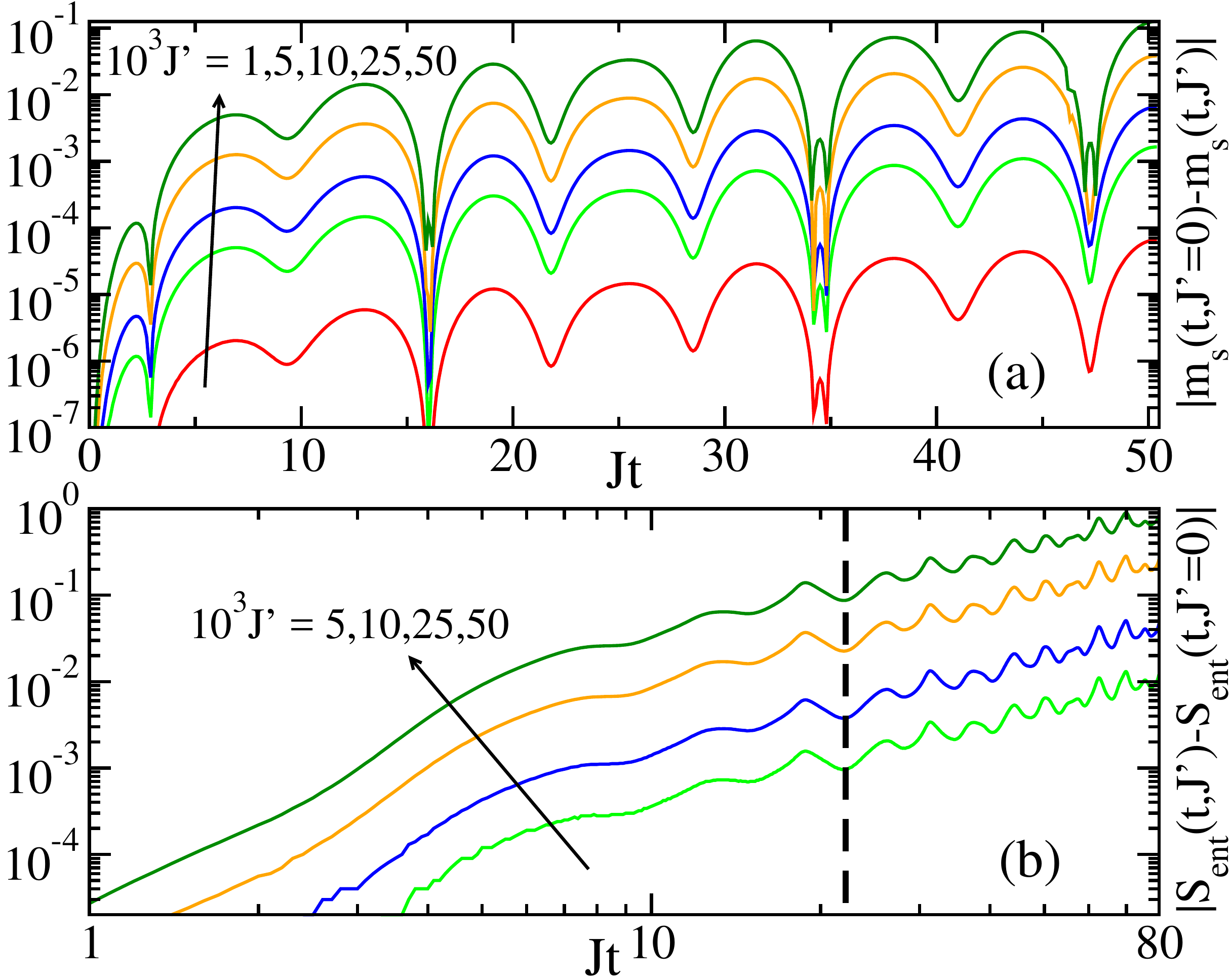}
\caption{(a) Difference between $m_s(t)$ for finite $J'$ and $m_s(t)$ for $J'=0$. After a rapid increase an exponential decay sets in for $Jt\gtrsim 5$. (b) Difference between $S_{\textrm{ent}}(t)$ for zero and finite $J'$. A power-law scaling with a small exponent $\alpha$ sets in for $Jt\gtrsim 20$ (dashed line). In both cases the onset of scaling is independent of $J'$.}
\label{Fig3b}
\end{figure}

One of the main observations in Ref.~\onlinecite{YaoLaumann}, where
the model \eqref{Ham} was studied for $\Delta=0$ on small clusters of
up to $L=8$ sites, was that in the dynamics distinct time scales are
visible, in particular, in the time evolution of the entanglement
entropy. More precisely, the study indentified the time scales $1/J$,
$1/J'$, $\e^L/J'$, and $D/(J')^2$. The latter two time scales are
irrelevant in our study because we are in the thermodynamic limit and
restrict ourselves to effectively infinite $D$. In the remaining
regimes the authors identified the following behavior for
$S_{\textrm{ent}}(t)$: (i) $0<t<1/J$: initial growth, (ii)
$1/J<t<1/J'$: single-particle localized plateau, and (iii) $t>1/J'$: a
logarithmic growth. In App.~\ref{App_A} we show that the same time
regimes are also visible for the initial state chosen in our study if
we consider the same $L=8$ cluster. The differences between
Ref.~\onlinecite{YaoLaumann} and our LCRG results are {\it not} a
result of different initial conditions but rather of the different
regimes investigated ($L\ll t$ versus the thermodynamic limit). In
particular, in the thermodynamic limit results for $m_s(t)$ in
Fig.~\ref{Fig1} and $S_{\textrm{ent}}(t)$ in Fig.~\ref{Fig3} no
qualitative changes occur at the time scale $1/J'$. Instead $m_s(t)$
shows an exponential decay for all times $Jt\gtrsim 5$ while the
asympotic power-law scaling for $S_{\textrm{ent}}(t)$ sets in for
$Jt\gtrsim 20$ independent of $J'$. What does change as a function of
$J'$ in a smooth way are the decay rate $\gamma$ and the power-law
exponent $\alpha$ with {\it both} $\gamma,\alpha\to 0$ for $J'\to 0$.
To confirm this picture we have also calculated both quantities for
small $J'$ such that $1/J'\in [20,1000]$. The results are shown in
Fig.~\ref{Fig3b} and are fully consistent with an onset of scaling
independent of $J'$.
To summarize, we have obtained a picture which is very different from
the one described in Ref.~\onlinecite{YaoLaumann} for small clusters:
There is no single-particle localized plateau in $S_{\textrm{ent}}(t)$
for $1/J<t<1/J'$ in the thermodynamic limit. In fact, the scale $1/J'$
is not visible at all in our data. There is also no quasi MBL regime
for $1/J'<t\ll D/(J')^2$ with logarithmic entanglement growth. Instead,
we find a scaling regime which becomes established at $Jt\gtrsim 20$
independent of $J'$ in which $m_s(t)\sim
\e^{-\gamma t}$ and $S_{\textrm{ent}}(t)\sim t^\alpha$ with
$\alpha<1$ where both $\gamma,\alpha$ are smooth functions of $J'$. We
have argued that this behavior can be explained by barriers of size
$N$ and a related dephasing time which scales as $\tau\sim N$ while
the entanglement time across a barrier scales as
$\tau_{\textrm{ent}}\sim N^{1/\alpha}$. Thus we have identified a time
regime with sub-ballistic entanglement spreading different {\it both}
from normal ergodic as well as MBL behavior. The exact diagonalization
results presented in App.~\ref{App_A} clearly demonstrate that the
behavior found in Ref.~\onlinecite{YaoLaumann} is the result of
finite-size effects and does not describe the behavior in the
thermodynamic limit. Given that there is no quenched disorder, this
should not come as a complete surprise.

\section{The XXZ case}
\label{XXZ}
Next, we study the case where the spins $\vec{S}_j$ are interacting,
$0<\Delta\leq 1$ to obtain additional evidence for the mechanism
depicted in Fig~\ref{Fig2}. For $J'=0$ and $D\gtrsim 0.3$ the system
is then in an MBL phase where $m_s(t)$ does not decay and
$S_{\textrm{ent}}\sim\ln t$ \cite{EnssAndraschkoSirker}. Turning on
the hopping $J'$ is expected to drive a phase transition from the MBL
into an ergodic phase. 

\subsection{Weak and intermediate interactions}
\begin{figure}[!ht]
\includegraphics*[width=0.49\linewidth]{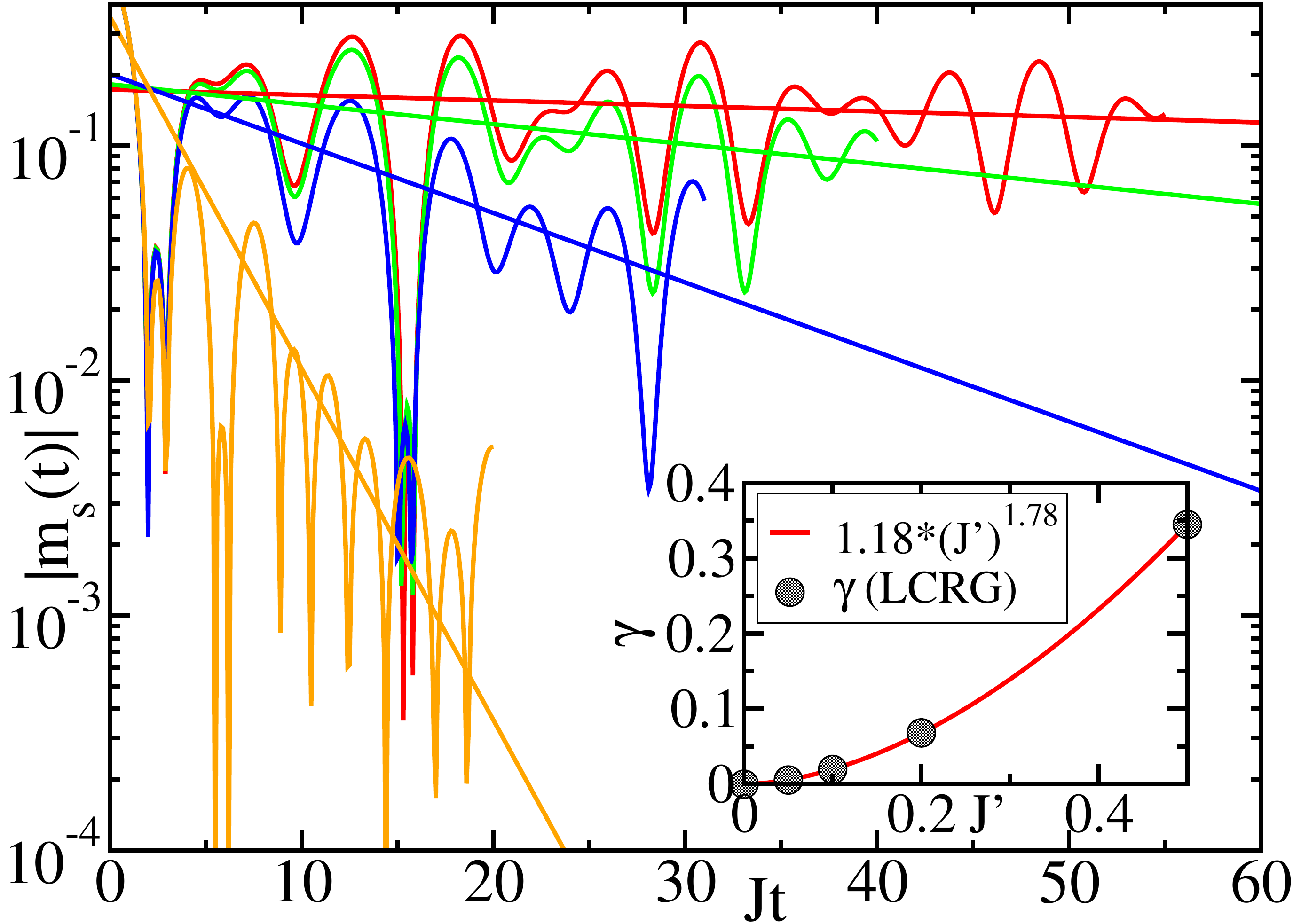}
\includegraphics*[width=0.49\linewidth]{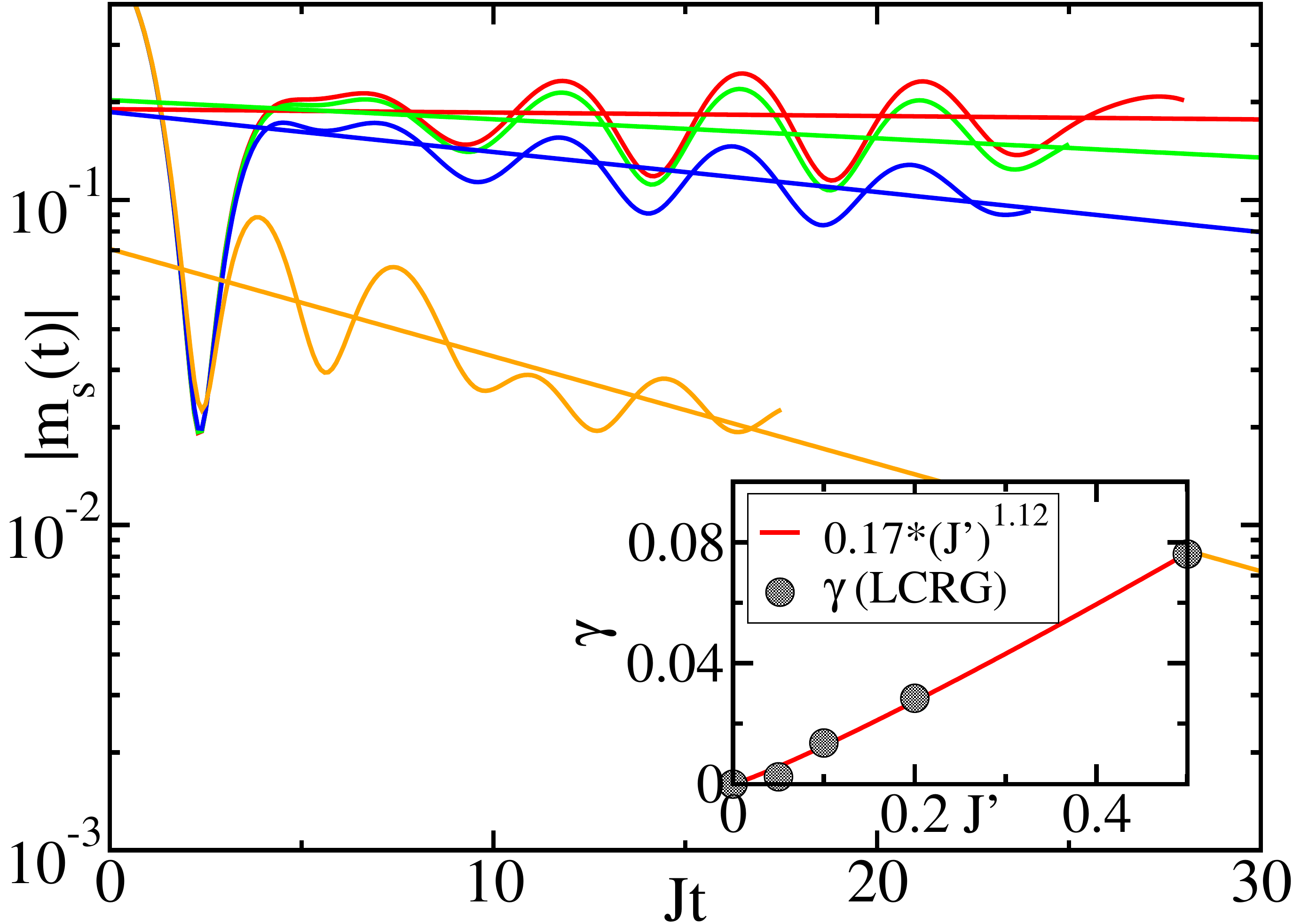}
\caption{$m_s(t)$ for $\Delta=0.2$ (left) and $\Delta=0.6$ (right). Curves are LCRG data and lines exponential fits. Insets: Decay rates as function of hopping $J'$.}
\label{Fig4}
\end{figure}

We start by analyzing the numerical data for $m_s(t)$ as a function of
$J'$ for $\Delta=0.2$ and $\Delta=0.6$, see Fig.~\ref{Fig4}.
While the oscillations in the magnetization curves are damped as
compared to the $\Delta=0$ case shown in Fig.~\ref{Fig1}, they still
show an exponential decay for $Jt\gtrsim 5$. The relaxation rate
$\gamma$ extracted from the fits increases monotonically with $J'$
with the exponent of the power-law scaling, $\gamma=A(J')^\beta$,
decreasing with increasing $\Delta$. We notice, furthermore, that
while the amplitude $A$ is similar to the XX case for $\Delta=0.2$, it
is an order of magnitude smaller for $\Delta=0.6$: At intermediate
interaction strengths we observe a pronounced slowing down of the
relaxation (see also Fig.~9).

\begin{figure}[!ht]
\begin{center}
\includegraphics*[width=0.498\linewidth]{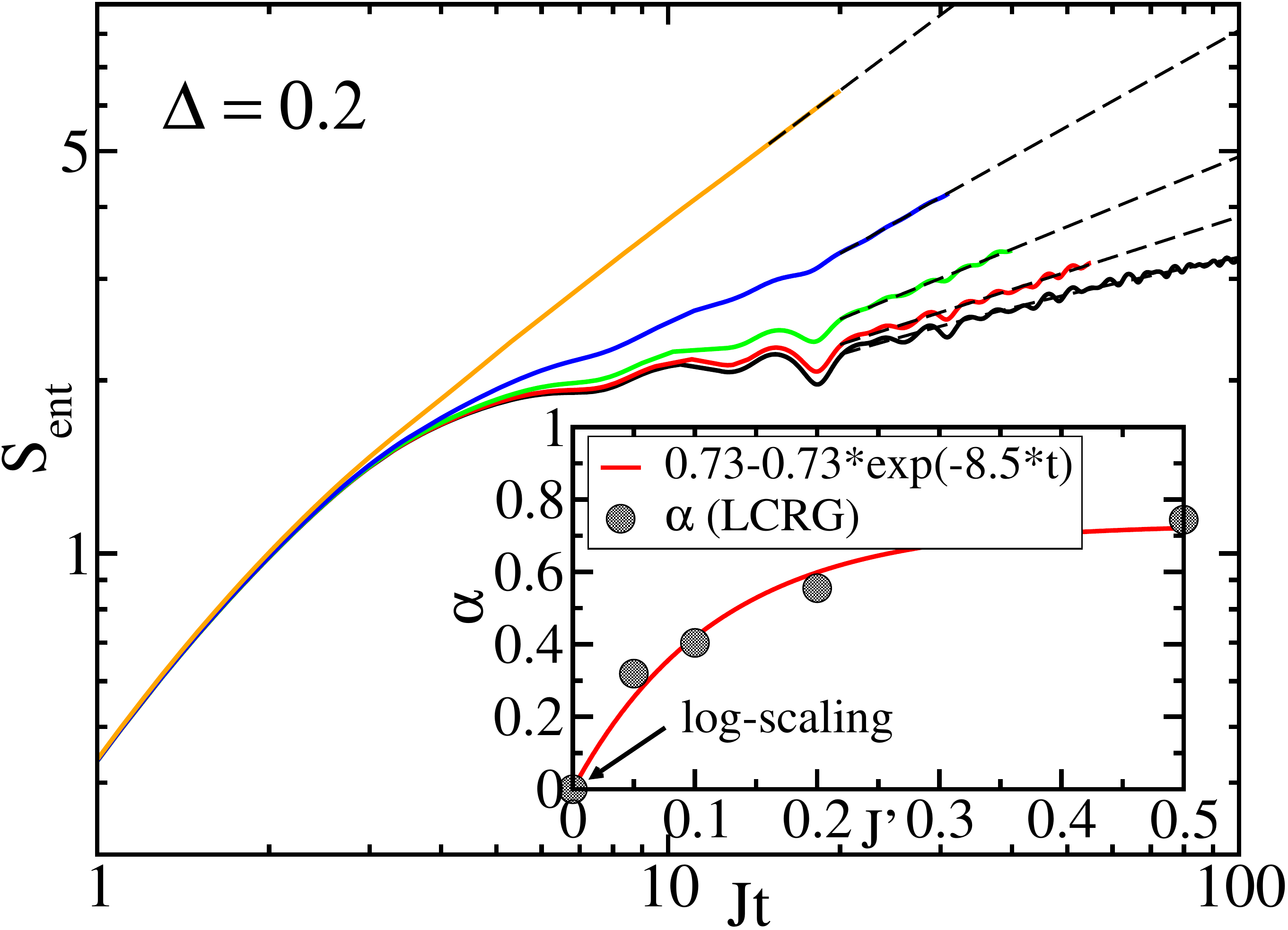}
\includegraphics*[width=0.488\linewidth]{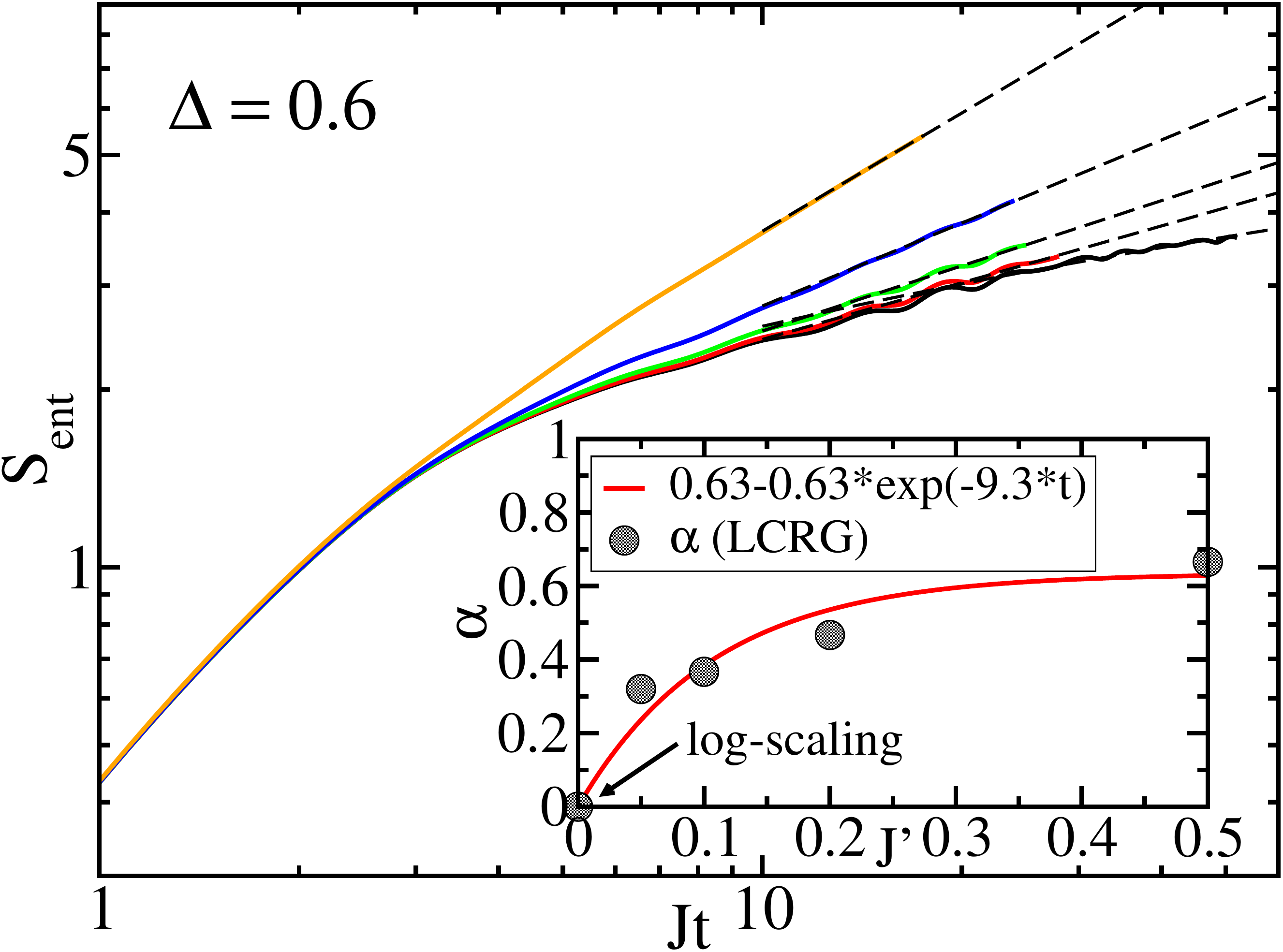}
\end{center}
\caption{Entanglement entropies for $\Delta=0.2$ and $\Delta=0.6$ and $J'=0,0.05,0.1,0.2,0.5$ (from bottom to top). The behavior is qualitatively similar to the $\Delta=0$ case shown in Fig.~\ref{Fig3} except for $J'=0$ where $S_{\textrm{ent}}\sim \ln t$.}
\label{Fig5}
\end{figure}

The entanglement entropy, shown in Fig.~\ref{Fig5}, also shows a
behavior very similar to the $\Delta=0$ case. The only qualitative
difference is the logarithmic scaling for $J'=0$ indicative of the MBL
phase.
The fitted exponents $\alpha$ of a power-law scaling (see inset of
Fig.~\ref{Fig5}) are again consistent with a sub-ballistic power-law
scaling $S_{\textrm{ent}}(t)\sim t^\alpha$ with $\alpha\approx 0.73$
for large $J'$.

In summary, interactions $\Delta$ reduce the relaxation rate
$\gamma$. The initial N\'eel state decays slower. Considering the
dephasing process shown in Fig.~\ref{Fig2} this is not surprising. The
flips of the spins $\vec{S}_j$---required to make the barrier formed
by the spins $\vec{\sigma}_j$ mobile---lead to a local ferromagnetic
arrangement. For finite $\Delta$ this now involves an energy cost. The
numerical results nevertheless show that the dephasing time across the
barrier has the same functional form given by $\tau=f(J',D,\Delta)N$
with a function $f(J',D,\Delta)$ which is growing with increasing
$\Delta$. The sub-ballistic scaling of the entanglement entropy is
qualitatively not affected by the interaction.

\subsection{The Heisenberg case}
\begin{figure}[!ht]
\includegraphics*[width=0.99\linewidth]{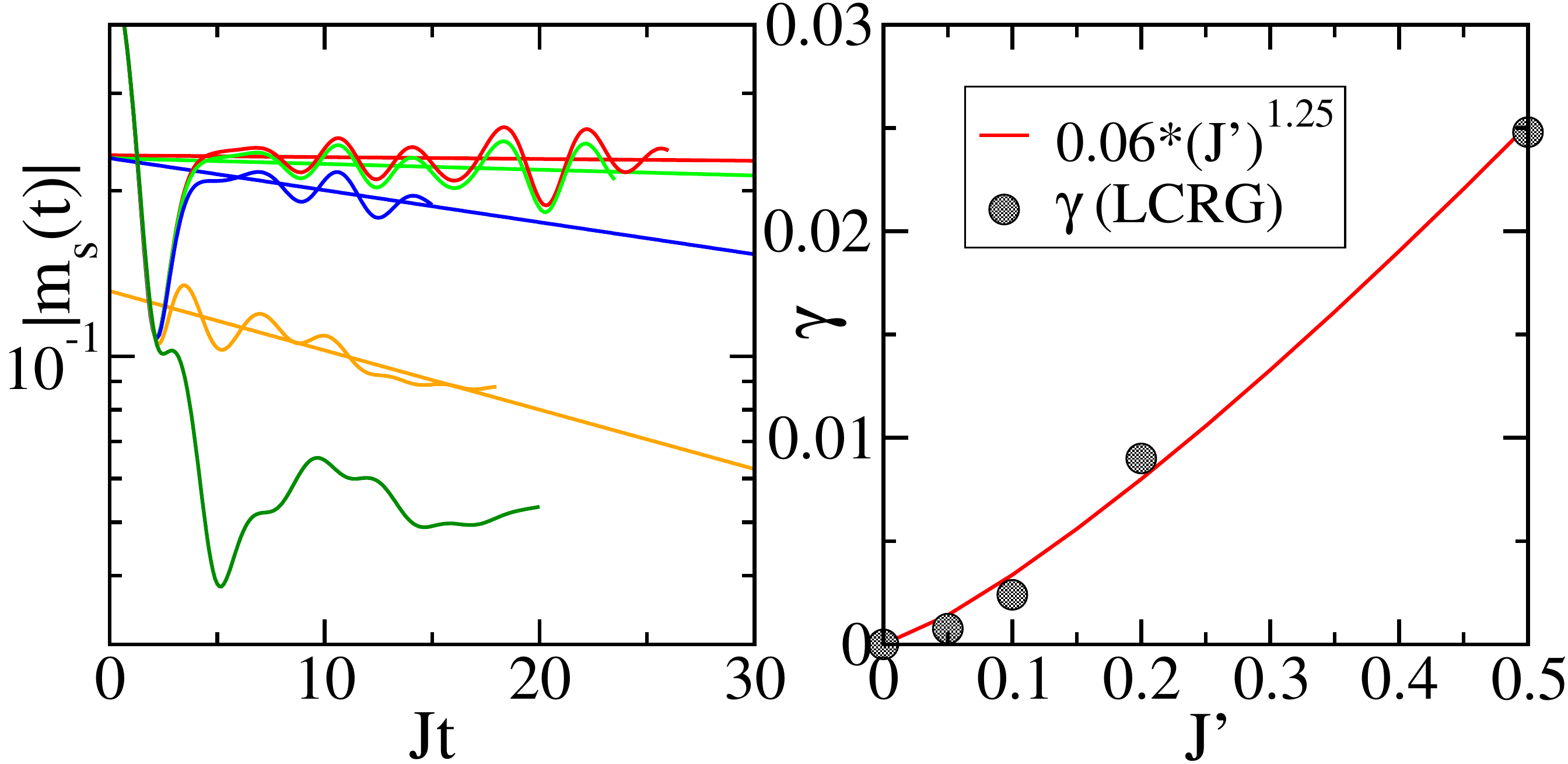}
\caption{$m_s(t)$ for $\Delta=1.0$ and $J'=0.05,0.1,0.2,0.5,0.7$ (left panel, from top to bottom). The data are consistent with a very slow exponential decay with relaxation rates $\gamma$ (right panel) except for $J'=0.7$ where in the accessible time range no clear scaling is observed.}
\label{Fig6}
\end{figure}

Finally, we also want to investigate the case of an SU(2) symmetric
exchange on one of the legs, i.e.~the case $\Delta=1$ in
Eq.~\eqref{Ham}. For the staggered magnetization we still find an
exponential decay for small $J'$, however, the relaxation rate is now
extremely small (see Fig.~\ref{Fig6}).
Furthermore, the data for $J'=0.7$ show large irregular oscillations
and no clear scaling. A likely explanation is that the numerically
accessible time scales are just too limited in this case. For the
couplings $J'\in [0,0.5]$, on the other hand, the relaxation rate
again shows a power-law scaling with a similar exponent as for
$\Delta=0.6$ but with an amplitude which is reduced by another factor
of $3$. This is consistent with the process depicted in Fig.~\ref{Fig2}.

\begin{figure}[!ht]
\begin{center}
\includegraphics*[width=0.99\linewidth]{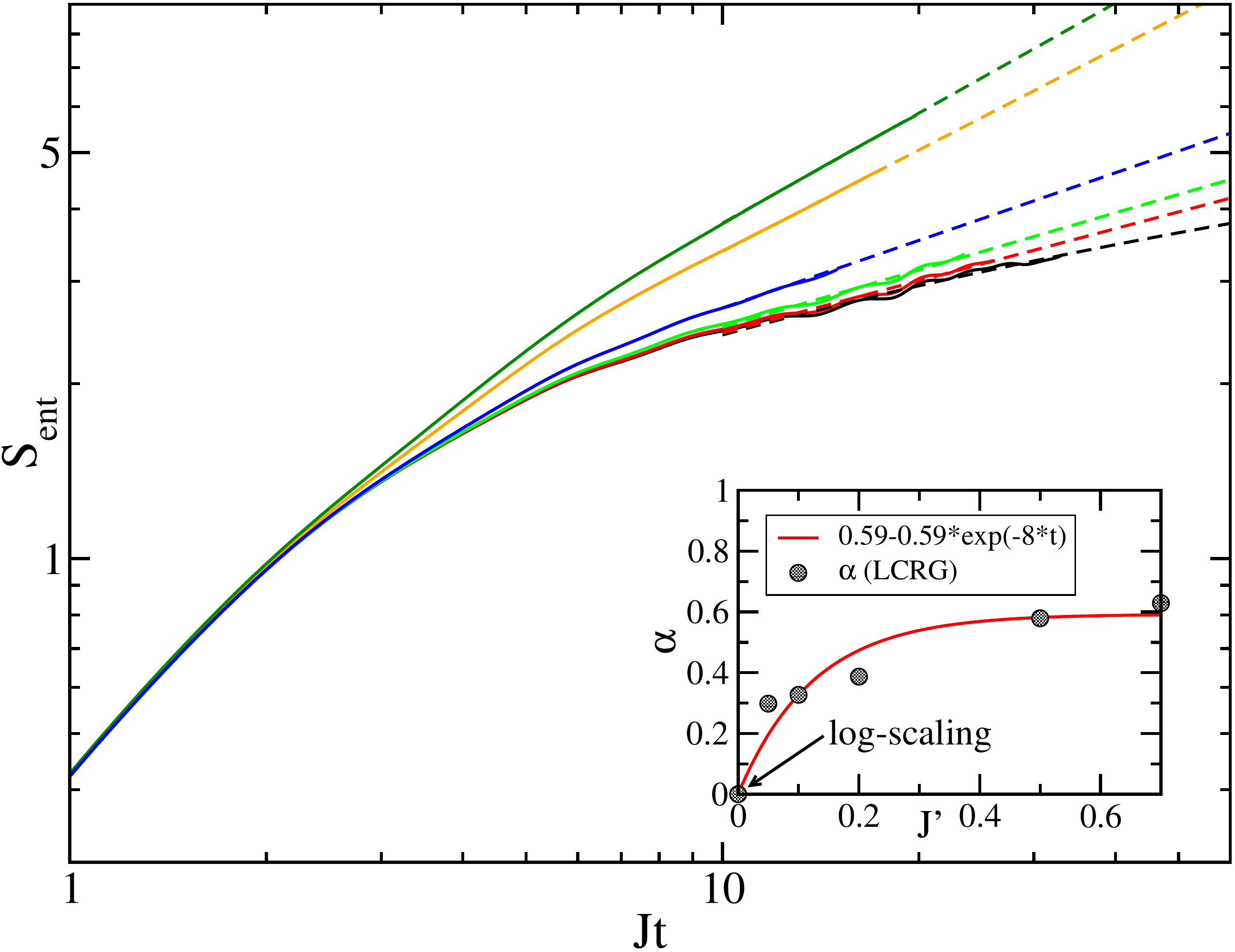}
\end{center}
\caption{Entanglement entropy for $\Delta=1.0$ and $J'=0,0.05,0.1,0.2,0.5,0.7$ (from bottom to top). Inset: Exponent $\alpha$ as a function of $J'$.}
\label{Fig7}
\end{figure}

The entanglement entropy changes very little as compared to the
$\Delta=0.6$ case (see Fig.~\ref{Fig7}).
Power-law fits still describe the data well for $Jt\gtrsim 10$. The
onset of the power-law scaling remains independent of $J'$. For the
extracted exponent $\alpha$ the behavior as function of $J'$ is still
the same as for the smaller $\Delta$ values, although the data are a
bit more scattered around the fit function. This is likely a
consequence of the limited times accessible numerically for
$\Delta=1$: The fits are more affected by oscillations on short time
scales and thus less reliable. Finally, we want to remark that there
is no extended regime in time where the entanglement growth is
independent of $J'$. This is contrary to the results for small
clusters discussed in Ref.~\onlinecite{YaoLaumann} showing again that
the results for such small system sizes are not indicative of the
behavior in the thermodynamic limit but are rather dominated by finite
size effects.

\section{Conclusions}
\label{Concl}
We have investigated quench dynamics in a spin ladder with a large
coupling $\sim DS^z_j\sigma^z_j$ along the rungs which is effectively
infinite in the time regime investigated. While the spins $\vec{S}_j$
with exchange amplitude $J$ and interaction $\Delta$ were prepared in
a N\'eel state, the spins in the other leg with exchange amplitude
$J'$ were prepared in a product state $\bigotimes_j (|+\rangle +
|-\rangle)_j/\sqrt{2}$. For $J'=0$ the spins $\vec{\sigma}_j$ realize
a quenched binary disorder potential for the spins $\vec{S}_j$. In
this case, the model is either in an AL phase ($\Delta=0$) or an MBL
phase ($\Delta\neq 0$).

Using an infinite-size density-matrix renormalization group algorithm
we have addressed the question whether the MBL phase can survive for
finite $J'$ ('MBL without disorder') or if there is an extended regime
in time where the model for finite $J'$ still shows the
characteristics of the MBL phase ('quasi MBL') including a logarithmic
increase of the entanglement entropy and memory of the initial
state. To answer these questions we have investigated the N\'eel order
parameter $m_s(t)$ and the entanglement entropy $S_{\textrm{ent}}(t)$
between two semi-infinite halfs of the ladder as a function of time
$t$ after the quench. The results clearly show that there is no MBL
phase for finite $J'$ and also no time regime where MBL
characteristics persist. The time regime $1/J<t\ll D/J^2$ for
$J'\neq 0$ is instead characterized by an exponential polarization decay
$m_s(t)\sim\e^{-\gamma t}$ with decay rate
$\gamma=\gamma(J',D,\Delta)$ while the entanglement entropy grows
sub-ballistically, $S_{\textrm{ent}}(t)\sim t^\alpha$ with
$\alpha<1$. The latter behavior is ``in between'' the logarithmic
growth in the MBL phase and the linear growth expected when the system
becomes fully ergodic.

\begin{figure}[!ht]
\begin{center}
\includegraphics*[width=0.75\linewidth]{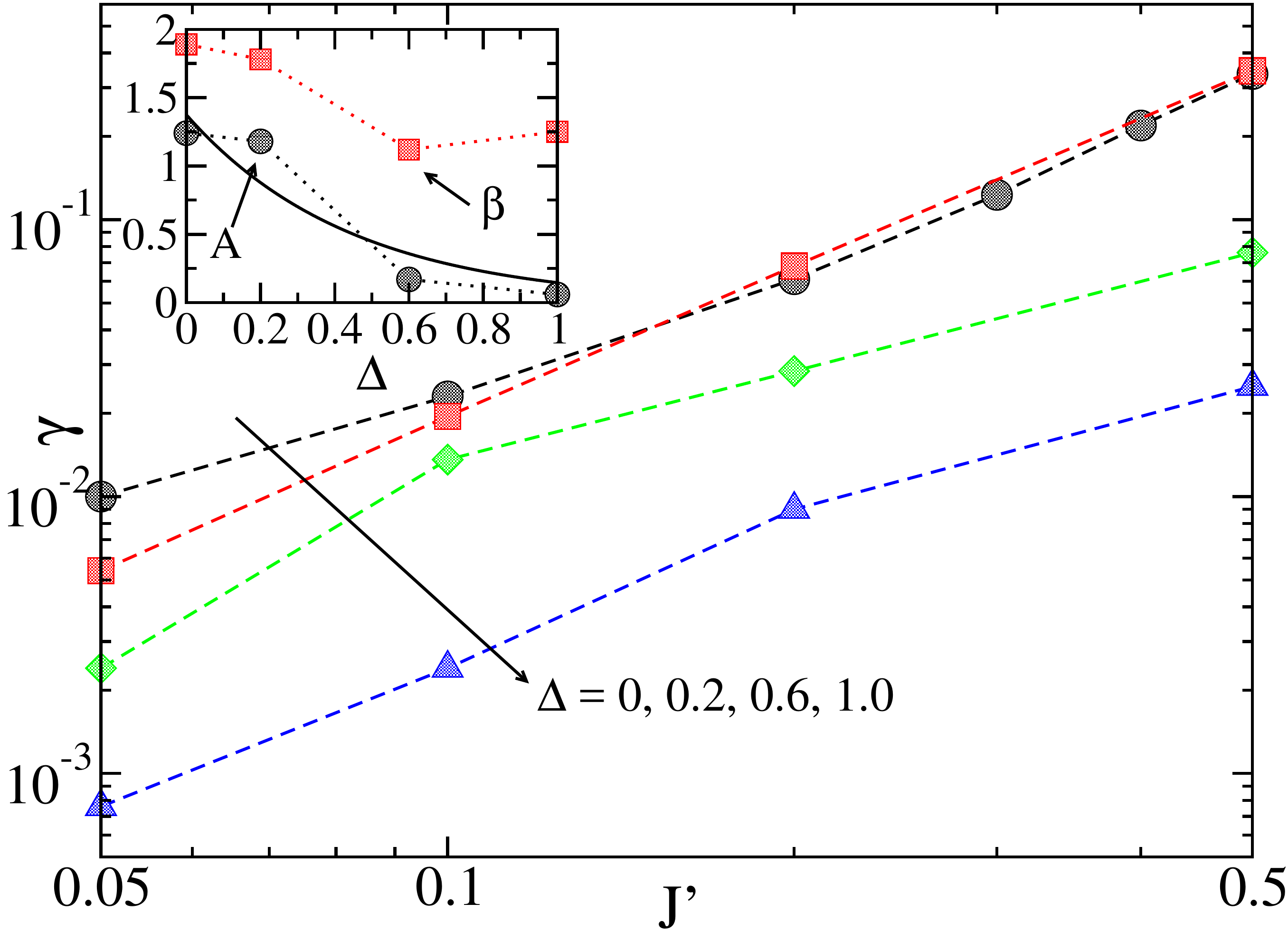}
\end{center}
\caption{Relaxation rates $\gamma$ extracted from the decay of $m_s(t)$  as a function of $J'$ for various $\Delta$. The results are consistent---with some larger deviations for $\Delta=0.6$---with a power-law scaling. Inset: Parameters $A,\beta$ obtained from fits $\gamma=A\, (J')^\beta$. The amplitude $A$ is well fitted by $A(\Delta)=1.4\cdot\exp(-2.2\Delta)$ (solid line).}
\label{Fig_relax_rates}
\end{figure}

An analysis of the numerical results for the decay rate of the
staggered magnetization for different $J'$ and $\Delta$ is consistent
with
\begin{equation}
\label{relax_rate}
\gamma \propto \text{e}^{-\Delta} J'^\beta 
\end{equation}
where $\beta \in [1,2]$ is a weakly $\Delta$-dependent exponent while
the amplitude of the power-law scaling is exponentially surpressed
with increasing $\Delta$, see Fig.~\ref{Fig_relax_rates}.
These results---including the exponential suppression of the amplitude with increasing $\Delta$---is consistent with the mechanism for the staggered magnetization decay explained and depicted in Fig.~\ref{Fig2}.

Our results are quite different from those obtained in a previous
exact diagonalization study, Ref.~\onlinecite{YaoLaumann}, which
concentrated mostly on the $\Delta=0$ case. In the latter study
distinct time regimes including a quasi MBL phase in time were
identified. None of this is confirmed in our study for infinite-size
systems. The comparison with exact diagonalization data presented in
App.~\ref{App_A} clearly show that the plateaus in
$S_{\textrm{ent}}(t)$ are a consequence of finite-size effects and are
not present in the thermodynamic limit.

In conclusion, we studied a prototypical model which has been put
forward as showing MBL without disorder or quasi MBL behavior and
concluded that neither one is realized. Our study, however, does not
exclude that different models exist which do show quasi MBL regimes or
where a power-law scaling of the entropy with a small exponent is
difficult to distinguish from a logarithmic
growth.\cite{MichailidisZnidaric}

\acknowledgments
The author thanks T.~Enss for discussions at an early stage of the
project. We acknowledge support by the Natural Sciences and
Engineering Research Council (NSERC, Canada) and by the Deutsche
Forschungsgemeinschaft (DFG) via Research Unit FOR 2316. We are
grateful for the computing resources provided by Compute Canada and
Westgrid as well as for the GPU unit made available by NVIDIA.

\appendix
\section{Exact diagonalization results}
\label{App_A}
\begin{figure}[!ht]
\begin{center}
\includegraphics*[width=0.99\linewidth]{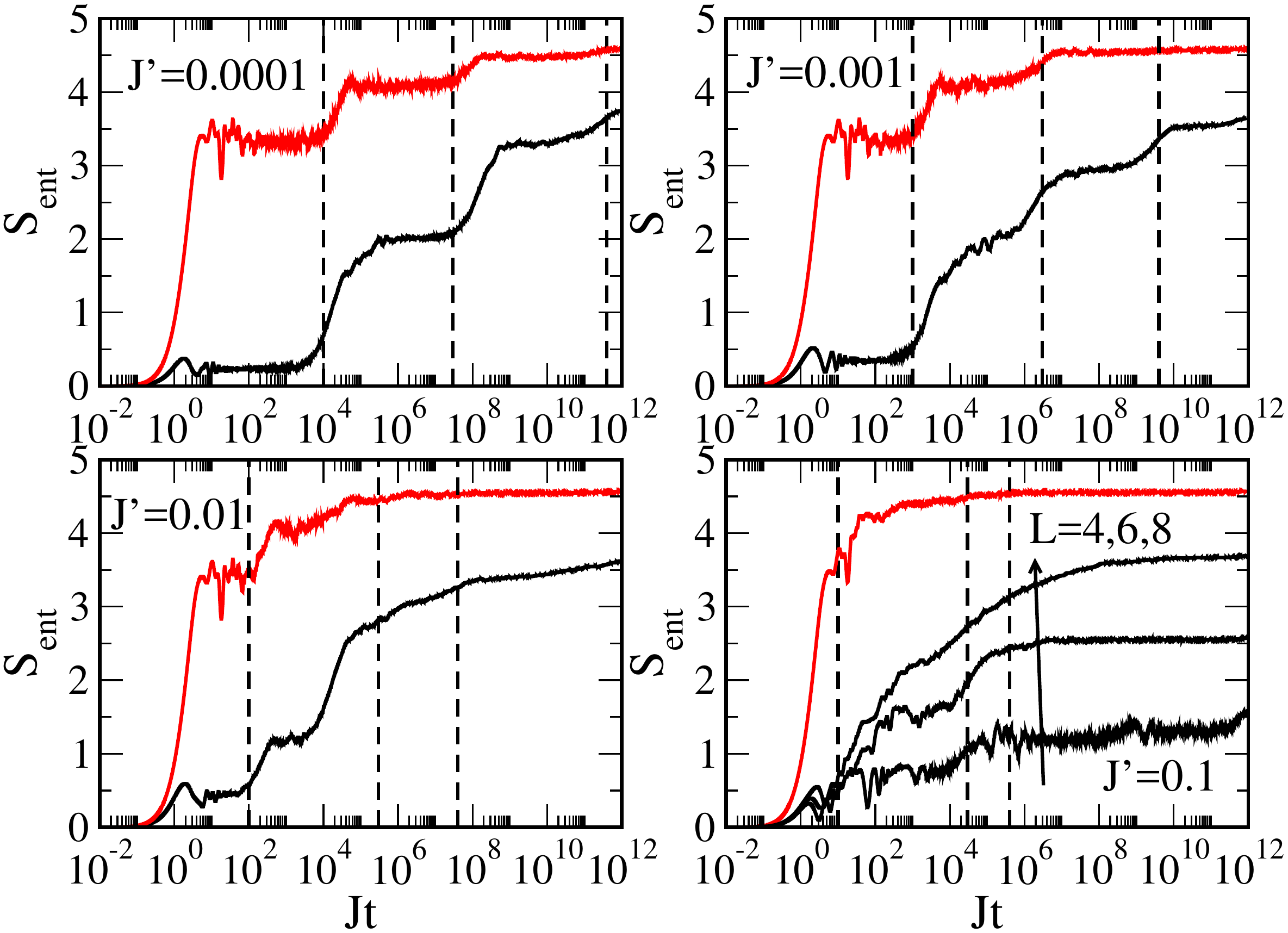}
\end{center}
\caption{Entanglement entropy during time evolution with the Hamiltonian \eqref{Ham} for $D=4000$ and $\Delta=0$ and different $J'$ for a $L=8$ ladder. The black lines are results for an average over $100$ initial product states, the red lines the result for the initial state $|\Psi\rangle = |N\rangle_S \otimes
|\infty\rangle_\sigma$. Dashed vertical lines denote the energy scales $1/J',\; \e^L/J',\; D/(J')^2$ (from left to right). For $J'=0.1$, data for $L=4,6$ are shown in addition. The time step in the numerical data is $\textrm{log}_{10}(t_{n+1}/t_n)=0.001$ and running averages over 100 time steps are shown for clarity.} 
\label{Fig_A1}
\end{figure}

The model \eqref{Ham} for $\Delta=0$ is exactly the same model which
has been studied in Ref.~\onlinecite{YaoLaumann} by exact
diagonalizations of ladders of up to $L=8$ sites. A difference between
the previous study and the results presented here for infinite systems
size are the initial states. While in the former case results for
averages over $30-100$ initial product states were presented with the
constraint that the same number of up and down spins are present on
each leg, we have considered a single initial state in which the spins
on one leg were prepared in the N\'eel state and the spins on the
other leg in the state $|\infty\rangle_\sigma =\bigotimes_j
\frac{1}{\sqrt{2}} (|+\rangle + |-\rangle)_j$. Given that the system does not contain quenched 
disorder, the precise initial state---as long as it is
non-trivial---should not {\it qualitatively} affect the dynamics at
times $t>1/J'$. This is supported by exact diagonalization data for
ladders with $L=8$ sites ($N=16$ total lattice sites) shown in
Fig.~\ref{Fig_A1}. Here the dynamics obtained by taking averages over
initial product states as in Ref.~\onlinecite{YaoLaumann} is compared
with dynamics starting from the initial state $|\Psi\rangle =
|N\rangle_S \otimes |\infty\rangle_\sigma$ as used in this study.

Importantly, the results for small $J'$ are qualitatively the same for
both sets of initial states. An initial rapid increase up to time
$\sim 1/J$ is followed by a first plateau which stretches out to $\sim
1/J'$. The entanglement entropy then increases again and reaches
another plateau which lasts up to the finite size scale $\sim
\e^L/J'$. This is followed by another increase and another plateau
extending up to the scale $D/(J')^2$. Our exact diagonalization
results thus confirm the results found in Ref.~\onlinecite{YaoLaumann}
for small clusters. For larger $J'$ these structures are somewhat less
visible, in particular, if we start in the initial state $|\Psi\rangle
= |N\rangle_S \otimes |\infty\rangle_\sigma$ because the entanglement
is larger and is limited by the maximally possible entanglement for
this cluster size ($S_{\textrm{ent}}^{\textrm{max}}=8\ln 2\approx
5.5$).

From the data for different system sizes obtained by averaging over
initial product states with $J'=0.1$ shown in the lower right panel of
Fig.~\ref{Fig_A1} it becomes clear that the plateaus in the entanglement 
entropy are the result of the finite-size structure of the
spectrum: For a total number of sites of the ladder $N=2L > 1/J'$
they completely disappear.

\begin{figure}[!ht]
\begin{center}
\includegraphics*[width=0.99\linewidth]{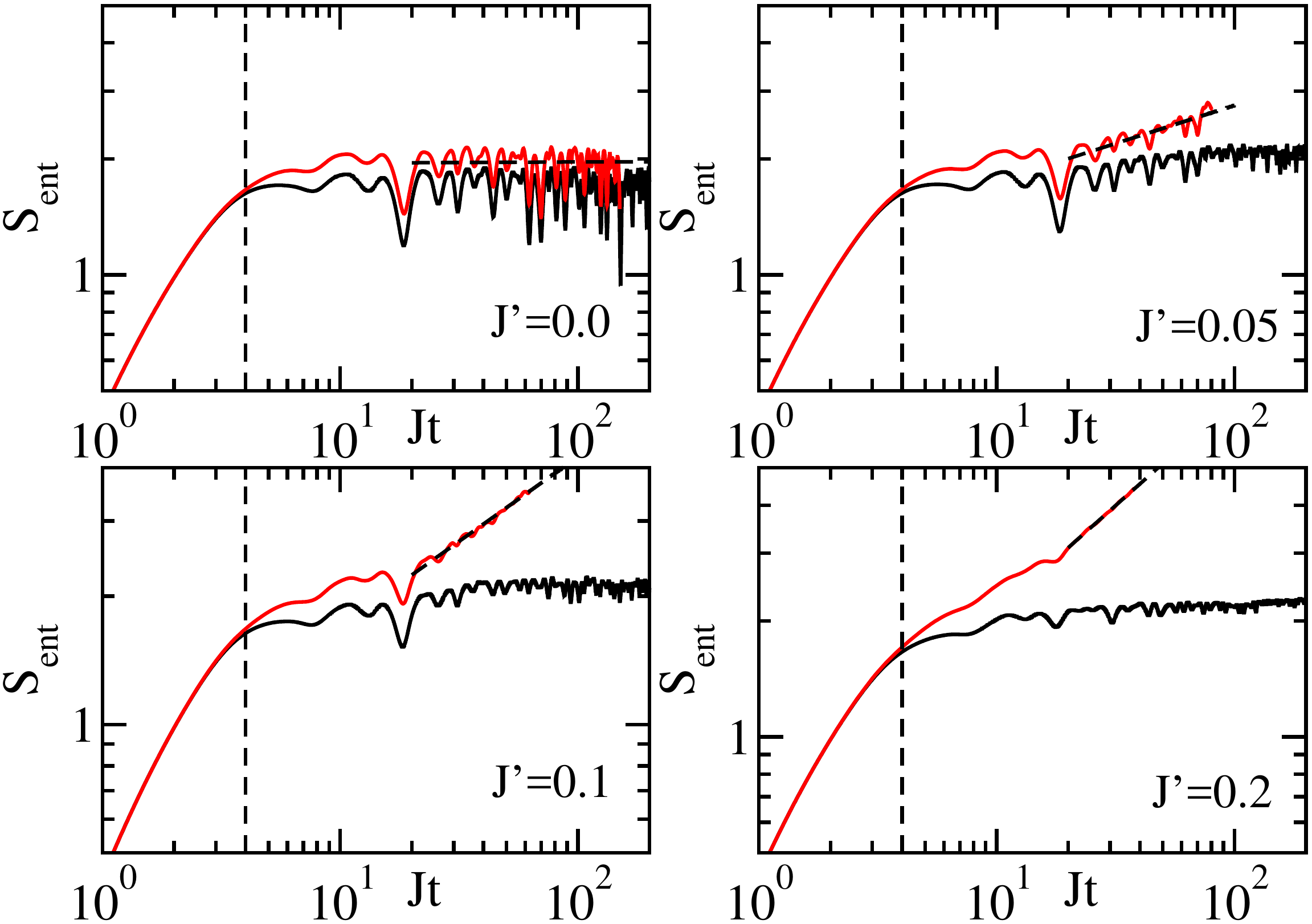}
\end{center}
\caption{$S_{\textrm{ent}}(t)$ during time evolution with the Hamiltonian \eqref{Ham} starting from the initial state $|\Psi\rangle = |N\rangle_S \otimes
|\infty\rangle_\sigma$ for $D=4000$, $\Delta=0$ and different $J'$. Compared are exact diagonalizations for $L=8$ ladders (black lines) with LCRG data in the thermodynamic limit (red lines with dashed lines representing power-law fits). Both only agree for $t\lesssim 1/J$ (dashed vertical line) due to finite size effects.} 
\label{Fig_A2}
\end{figure}

Finally, we want to demonstrate that the exact diagonalization data
cannot be used to infer the behavior of the system in the
thermodynamic limit after the first rapid increase of
$S_{\textrm{ent}}(t)$, i.e. for times $t>1/J$. The direct comparison
of the LCRG data in the thermodynamic limit with exact diagonalization
data for the initial state $|\Psi\rangle = |N\rangle_S \otimes
|\infty\rangle_\sigma$ presented in Fig.~\ref{Fig_A2} makes it clear
that the generic power-law increase of the entanglement entropy for
times $t\gtrsim 1/J$ and $J'\neq 0$ is missed in exact
diagonalizations which instead show a plateau because of finite size
effects.


\end{document}